# Sub-Wavelength Terahertz Spin-Flip Laser Based on a Magnetic Point-Contact Array

R. I. Shekhter,[1] A. M. Kadigrobov,[1,2,a)] M. Jonson,[1,3,4] E. I. Smotrova,[5]
A. I. Nosich,[5] and V. Korenivski[6]

[1]*Department of Physics, University of Gothenburg, SE-412 96 Göteborg, Sweden*
[2]*Theoretische Physik III, Ruhr-Universität Bochum, D-44801 Bochum, Germany*
[3]*SUPA, Department of Physics, Heriot-Watt University, Edinburgh EH14 4AS, Scotland, UK*
[4]*Division of Quantum Phases and Devices, School of Physics, Konkuk University, Seoul 143-701, Korea*
[5]*Institute of Radiophysics and Electronics, National Academy of Sciences of Ukraine, Kharkiv 61085, Ukraine*
[6]*Nanostructure Physics, Royal Institute of Technology, SE-106 91 Stockholm, Sweden*

**Abstract.** We present a novel design for a single-mode, truly sub-wavelength THz disk laser based on a nano-composite gain medium comprising an array of metal/ferromagnetic point contacts embedded in a thin dielectric layer. Stimulated emission of light occurs in the point contacts as a result of spin-flip relaxation of spin-polarized electrons that are injected from the ferromagnetic side of the contacts. Ultra-high electrical current densities in the contacts and a dielectric material with a large refractive index, neither condition being achievable in conventional semiconductor media, allows the thresholds of lasing to be overcome for the lowest-order modes of the disk, hence making single-mode operation possible.

[a)] Electronic mail: anatoli.kadygrobov @physics.gu.se

The prospect of combining electronics and photonics on a single chip by integrating lasers with electronic devices in high-density integrated circuits is an exciting one. This would, however, require the on-chip lasers to be small and operate in the sub-millimeter range of wavelengths, which corresponds to the terahertz frequencies (300 GHz to 10 THz) normally associated with vacuum-tube and solid-state electronic oscillators. Nevertheless, for such relatively low frequencies (compared to optical microresonator lasers) it has already been possible to fabricate disk lasers with diameters ($2a$) somewhat smaller than the emission wavelength $\lambda$, with $2a/\lambda = 0.7$ for the smallest disks.[1,2] This was achieved by cascading 200 and more quantum wells in a semiconductor disk, which lead to a gain of $g = 27\ \text{cm}^{-1}$ and made it possible to overcome the lasing threshold for the mode with azimuthal index $m$ = 5. However, a further miniaturization of laser devices employing dielectric resonators hits a limit. This is because of the relatively low gain that can be achieved with typical active media such as semiconductors, dyes in polymer matrices, and ion-doped crystalline materials.

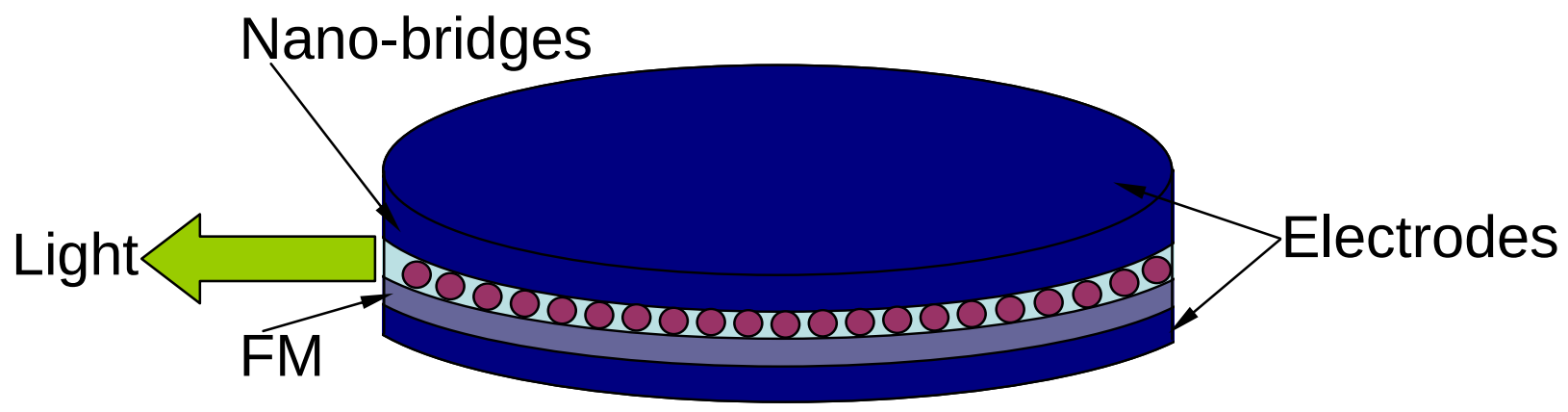

FIG. 1. (Color online) Configuration of the proposed disk laser containing a novel nano-composite active material comprising an ensemble of nano-bridges (point contacts) embedded in a thin dielectric layer.

In this Letter we propose to use a novel active medium to solve this problem. The idea is to use a nano-composite material (see Fig. 1) in which an ensemble of nano-bridges (point contacts), embedded in a thin dielectric with a large refractive index, connect a metallic and a ferromagnetic 2D layer to form a disk-shaped resonator. In such point contacts high-intensity light can be generated if spin-polarized electrons are injected from the magnetic material.[3,4] This is because an inverted population of electron energy levels, split by an external magnetic field or the exchange interaction in the material, leads to photon emission accompanied by spin-flip relaxation of the electrons. The intensity of the emitted radiation can be high due to the large, metallic density of the "hot" electrons that serve as photon emitters.

The spin flips affect the magneto-resistance of the point contact and have been predicted to give rise to a characteristic peak in the resistance for a magnetic field that corresponds to resonant electron-photon interactions, a prediction that has been confirmed by recent experiments.[4] As shown in Fig. 2, a one-parameter fit gives very good agreement between the theoretical prediction and the experiment, which demonstrates that the irradiated single point contact studied in Ref. 4 can be considered as an optically active medium with an optical gain of [5] $g_{pc}$ ~ 10-100 $cm^{-1}$. We suggest that an ensemble of such point contacts, embedded in a suitable dielectric material, could be used as the active medium in a lasing device.

The relaxation of the electrical injection current leads to Joule heating of the point contact, but on the other hand the electron flow carries heat away from the contact. To avoid overheating, small point contacts of linear size $d \approx 10^2$ nm are needed, with an area density of $n \approx (10d)^{-2}$. In this case the average optical gain, $g = 0.1 g_{pc}$ , can be estimated to be of order 0-10 $cm^{-1}$, depending on the voltage applied to the point contact.[6] For the active region (the point contact core) made of a normal-metal and subjected to a magnetic field of the order of 10 tesla, the frequency of the emitted light is approximately 0.3 THz, which is at the low-frequency limit of the THz range[7]. For ferromagnetic point contacts, the expected frequency range is determined by the exchange spin-splitting in the material, typically 1-30 THz, and can be controlled by, for example, changing the composition of the ferromagnet and thereby its Curie temperature.

We propose to build a THz laser as a thin multilayered disk whose central layer is a 10-20 nm dielectric sandwiched between normal-metal and ferromagnetic slabs which are connected by nano-bridges protruding through the dielectric as shown in Fig. 1. Such a metal-dielectric disk is an open resonator for which the disk diameter determines the working-mode type and hence the wavelength of the emitted light. Off the disk plane, electromagnetic-field confinement is provided by metallization of the faces of the disk; in the disk plane confinement is due to the different refractive indices of the inner and outer materials.

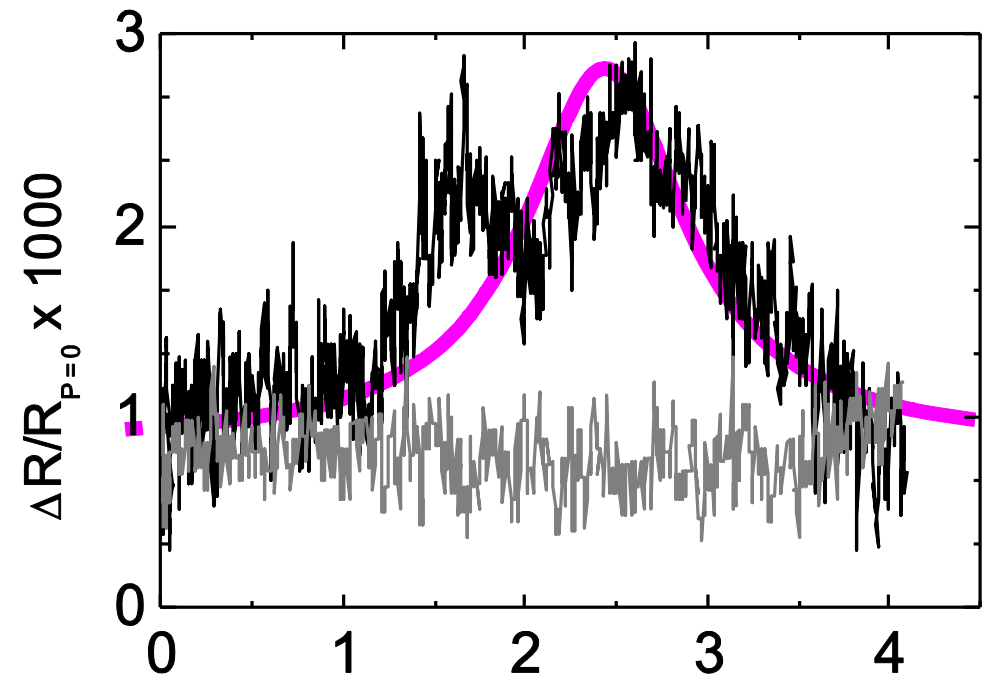

Fig.2: Magnetoresistance of Cu[100nm]/ $Fe_{0.7}Cr_{0.3}$(50nm) / Cu(3nm)-Cu point contact as a function of the external magnetic field. Fitting with the theoretical prediction is achieved by choosing the spin-flip relaxation frequency $\nu_{so} = 10^{11} s^{-1}$

It is convenient to study the lasing modes, including their wavelengths, thresholds and field patterns as a lasing eigenvalue problem (LEP), where Maxwell's equations are solved with appropriate boundary conditions on the disk surface and the outgoing-radiation condition at infinity. Full details of this newly developed approach and its application to various microlasers can be found in Refs. 8-10 (see also Refs. 11-15). The LEP eigenvalues are pairs of real numbers corresponding to the lasing frequency (or wavenumber, $k$) of the mode and the threshold value of material gain $\gamma$ (the imaginary part of the refractive index in the active region). Hence the eigenvalues give the spectrum and thresholds of the lasing radiation. The average optical gain per unit length, *g*, is connected to the material gain as $g = k\gamma$.

For the considered nano-composite disk whose top and bottom faces are metallic and whose thickness is a fraction of the wavelength, a useful approximation can be obtained if one assumes the metallic faces to be perfect conductors. Then the lasing modes can be viewed as TE-type modes in a 2-D circular dielectric uniformly active resonator. In this case the LEP is reduced, for each azimuthal order *m*, to a characteristic equation given, for instance, by Eq. (4) of Ref.8. Results of the LEP analysis for such a model of a disk resonator filled with material of refractive index $\alpha = 3.6$ (i.e., the same as in Ref. 2) are shown in Fig. 3. Discrete LEP solutions are shown as points in the plane $(2a/\lambda, \gamma)$. Note that while the thresholds obtained in this way correctly account for the radiation losses, they neglect the losses in the conductors. Hence the total losses are underestimated, which should be more important for higher-order modes.

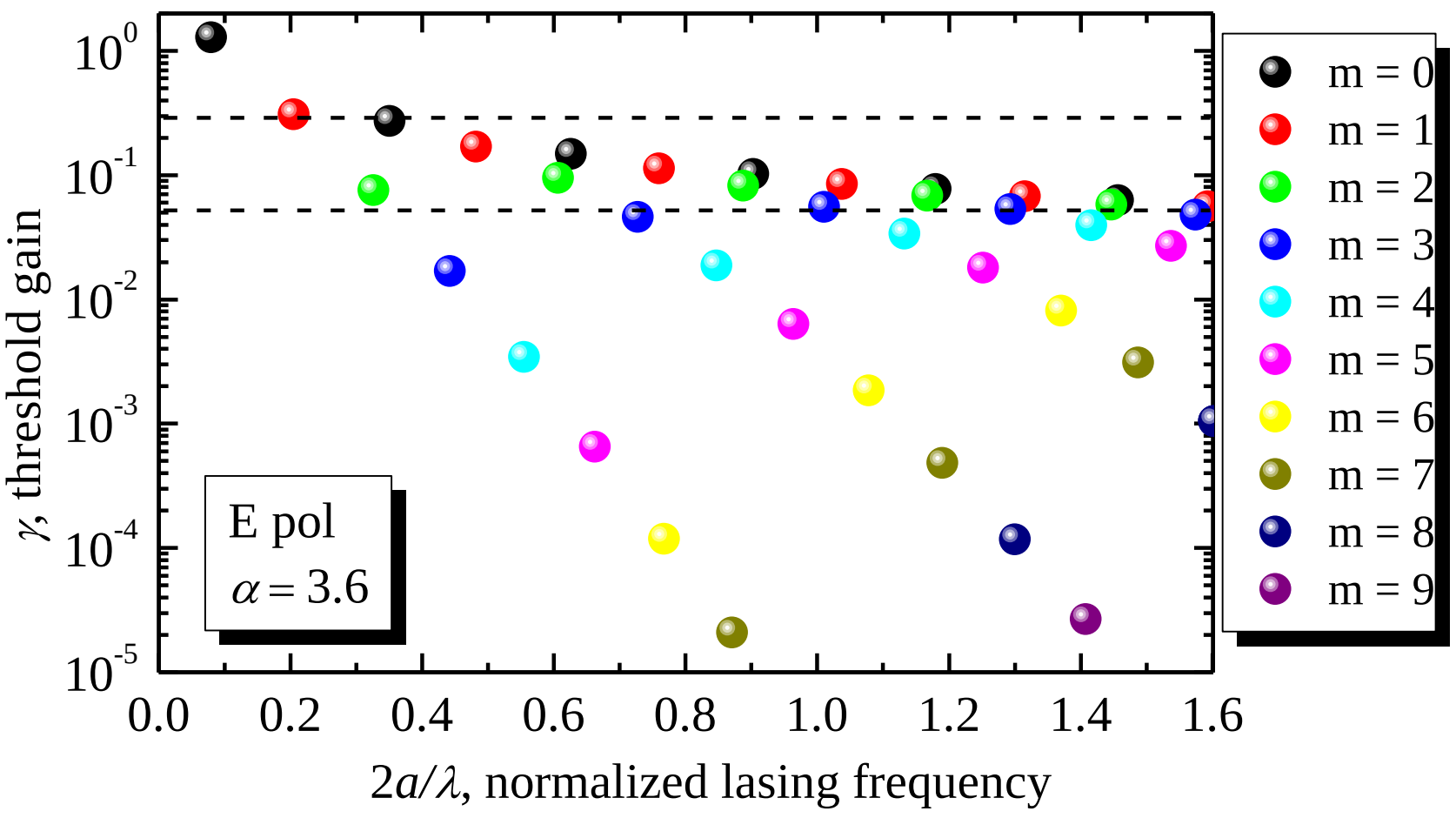

FIG. 3. (Color online) Lasing spectra and threshold gains for $TE_{mn}$ modes in a disk with bulk refractive index α = 3.74. Within one azimuthal family (same *m*) the radial index *n* takes the values 1, 2, 3,… from left to right. The upper dashed line indicates a gain of $\gamma = 0.3$ ($g = 160\ \text{cm}^{-1}$ at 3 THz), which can be realistically achieved with the proposed high-gain nano-composite medium. The lower dashed line at $\gamma = 0.043$ corresponds to the gain of 27 cm$^{-1}$ achieved at the same frequency in Ref. 2 with a semiconductor medium containing a cascade of 270 quantum wells.

As the refractive index varies from 1.5 to 10, Fig. 4-a shows that the characteristic frequencies of the lower-order E-type modes of the disk, $E_{01}$ and $E_{11}$, decrease and vary from $2a/\lambda \approx 0.2$ to 0.02. The fundamental mode $E_{01}$ is of particular interest since the electric field has a maximum in the center of the disk. However, because of its field pattern the threshold gain is considerably higher than for the other modes, including the $E_{11}$-mode. As one can see from Fig. 4-b, a laser operating in the $E_{11}$-mode needs to overcome a threshold gain ranging from 0.4 to 0.2 if the refractive index varies between 1.5 and 5. The necessary material-gain values appear obtainable in the proposed nano-composite medium, for which the material gain depends both on the voltage bias applied to the point contacts[6] and on the density of point contacts embedded in the dielectric inner domain of the sandwich-like thin metal-dielectric disk.

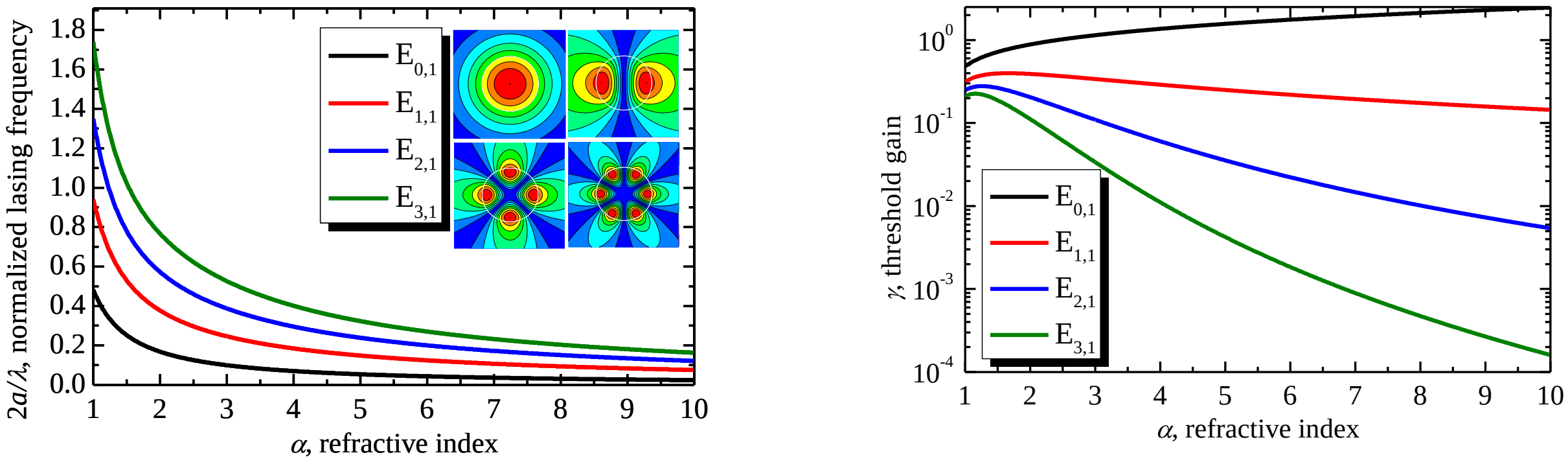

FIG. 4. (Color online) Normalized lasing frequencies (left panel) and threshold gains (right panel) for the $E_{01}$ , $E_{11}$, $E_{21}$ and $E_{31}$ modes plotted as a function of the resonator refractive index $\alpha$. Insets show the field patterns of the different modes at $\alpha = 3$.

It should be noted that the existing microdisk semiconductor lasers, which use whispering-gallery modes at optical and THz frequencies, are multi-mode devices. Several modes get into the emission range of the active material, compete with each other, and lead to mode hopping in the non-linear regime. Working in one of the lower-order modes makes laser operation more stable as the distance (in frequency) to the nearest parasitic mode is larger than in the case of whispering-gallery modes. For the case of normal metal point contact cores, an additional advantage of the proposed micro-disk laser is that it can be efficiently tuned by changing the applied magnetic field. Furthermore, unlike semiconductor microdisk lasers, a nano-composite disk laser can use novel dielectric materials with very high refractive indices, ranging from 5 ($HfO_2$) to the "colossal" value of 50, and reasonably low losses[16]. As simulations show (see. Fig. 4), if $\alpha = 5$, then even the modes $E_{21}$ and $E_{31}$ demonstrate the field confinement along the rim of the disk typical for whispering-gallery modes and correspondingly low thresholds, while the cavity size is deep inside the subwavelength range ($2a/\lambda < 0.3$).

Small-size cavities working in the lower-order modes have a small number (e.g, $2m = 2$ or 4) of lobes in the far-field radiation patterns. This is more convenient than the $2m >> 1$ identical beams typically emitted from lasers working with whispering-gallery modes. Besides, the thresholds for the lower modes are less sensitive to the precise shape of the disk boundary. Hence, the manufacturing tolerances can be relaxed. The near-field patterns for the four lowest modes in the disk resonator are shown as insets in Fig. 4.

In conclusion we have proposed a new design principle for a single-mode, truly sub-wavelength THz disk laser, which is based on using an active medium that allows the material gain to be several orders of magnitude higher than if conventional semiconductor media are used. This is because the proposed nano-composite active medium contains an array of metal-ferromagnetic nano-bridges (point contacts), from which light is emitted as a high-density, spin-polarized current relaxes in a process where the electrons flip their spin in a "vertical" transition between Zeeman- or exchange-split energy levels.

Financial support from the European Commission (FP7-ICT-FET No. 225955 STELE), the Swedish VR, the Korean WCU program funded by MEST/NFR (R31-2008-000-10057-0), and the Ukrainian NTNM target program (No. 1.1.3.45) is gratefully acknowledged.

1. G. Fasching, A. Benz, K. Unterrainer R. Zobl, A. M. Andrews, T. Roch, W. Schrenk, and G. Strasser**,** Appl. Phys. Lett. **87**, 1112 (2005).
2. G. Fasching, V. Tamosiunas, A. Benz, A. M. Andrews, K. Unterrainer, R. Zobl, T. Roch, W. Schrenk, and G. Strasser, IEEE J. Quantum Elect. **43**, 687 (2007).
3. A. Kadigrobov, Z. Ivanov, T. Claeson, R. I. Shekhter, and M. Jonson, Europhys. Lett. **67**, 948 (2004).
4. A. M. Kadigrobov, R. I. Shekhter, S. I. Kulinich, M. Jonson, O. P. Balkashin, V. V. Fisun, Yu. G. Naidyuk, I. K. Yanson, S. Andersson, and V Korenivski, New J. Phys. **13**, 023007 (2011).
5. If photons are emitted from a ferromagnetic part of the point contact the gain can be even higher. In this case a new mechanism for photon-induced electronic spin-flips based on the exchange interaction was predicted in Ref. 3, where it was shown that the optical gain can be larger by a factor $(c/v_F)^2$ and approach $10^5$-$10^6$ $cm^{-1}$.
6. The optical gain in a point contact of the type proposed can be expressed as[4] $g_{pc} = g_0 V / V^*$, where $g_0 = 3\pi^2 n_0 \mu^2{}_B \omega / \varepsilon_F \mathrm{v}_F$, $\varepsilon_F$ and $\mathrm{v}_F$ are the normal metal Fermi energy and Fermi velocity, $n_0$ is the electron density, $\mu_B$ is the Bohr magneton, $\omega$ is the radiation frequency, $V$ is the applied voltage, and $eV^* \sim \hbar\omega$.
7. For fields in the range 0.1-10 T the device would act as a MASER and emit microwaves in the range 3-300 GHz.

8. E. I. Smotrova, A. I. Nosich, T. M. Benson, and P. Sewell, IEEE J. Sel. Top. Quant. **11**, 1135 (2005).
9. A. I. Nosich, E. I. Smotrova, S.V. Boriskina, T. M. Benson, and P. Sewell, Opt. Quant. Electron. **39**, 1253 (2007).
10. E. I. Smotrova, V. O. Byelobrov, T. M. Benson, P. Sewell, J. Ctyroky, and A. I. Nosich, IEEE J. Quantum Elect. **47**, 20 (2011).
11. M. J. Noble, J. P. Loehr, and J. A. Lott, IEEE J. Quant. Electron. **34**, 1890 (1998).
12. J. V. Campenhout, P. Bienstman, and R. Baets, IEEE J. Select. Area. Comm. **25**, 1418 (2005).
13. X. Sun, J. Scheuer, and A. Yariv, IEEE J. Sel. Top. Quant. **13**, 359 (2007).
14. C. Manolatou and F. Rana, IEEE J. Quantum Elect. **44**, 435 (2008).
15. A. Mizrahi, V. Lomakin, B. A. Slutsky, M. P. Nezhad, L. Feng, and Y. Fainman, Opt. Lett. **33**, 1261 (2008).
16. S. Krohns, P. Lunkenheimer, Ch. Kant, A. V. Pronin, H. B. Brom, A. A. Nugroho, M. Diantoro, and A. Loidl, Appl. Phys. Lett. **94**, 122903 (2009).